\documentclass[a4paper,10pt,twoside]{cpc-hepnp}
\usepackage{CJK,upgreek,fancyhdr}
\usepackage{multicol}
\usepackage{booktabs}
\usepackage{amssymb,bm,mathrsfs,bbm,amscd}
\usepackage[tbtags]{amsmath}
\usepackage{lastpage}
\usepackage{authblk}
\usepackage{graphicx}
\usepackage{hyperref}
\usepackage{array}
\graphicspath{ {cpc-template-english/} }

\begin{document}



\title{Constraint on the Ground-state Mass of $^{21}$Al and Three-Nucleon Forces}

\author[1]{X.Z. Teng}
	\author[1]{Teh C.E.}
	\author[1]{J. Lee\thanks{jleehc@hku.hk}}
	\author[1,2]{X.X. Xu\thanks{xinxing@hku.hk}}
	\author[2,3]{C.J. Lin\thanks{cjlin@ciae.ac.cn}}
	\author[2]{L.J. Sun}
	\author[4]{J.S. Wang}
	\author[5]{D.Q. Fang}
	\author[1]{S. Leblond}
	\author[6]{Z.H. Li}
	\author[6]{J. Li}
	\author[2]{N.R. Ma}
	\author[4]{J.B. Ma}	
	\author[6]{H.L. Zang}
	\author[4]{P. Ma}
	\author[4]{S.L. Jin}
	\author[4]{M.R. Huang}
	\author[4]{Z. Bai}
	\author[1]{J.J. Liu}
	\author[1]{T. Lokotko}
	\author[2]{F. Yang}
	\author[2]{H.M. Jia}
	\author[2]{D.X. Wang}
	\author[4]{Y.Y. Yang}
	\author[4]{Z.G. Hu}
	\author[4]{M. Wang}
	\author[2]{H.Q. Zhang}

\affil[1]{Department of Physics, The University of Hong Kong, Hong Kong, China}
\affil[2]{China Institute of Atomic Energy, Beijing 102413, China}
\affil[3]{College of Physics and Technology, Guangxi Normal University, Guilin 541004, China}
\affil[4]{Institute of Modern Physics, Chinese Academy of Sciences, Lanzhou 730000, China}
\affil[5]{Shanghai Institute of Applied Physics, Chinese Academy of Sciences, Shanghai 201800, China}
\affil[6]{School of Physics and State Key Laboratory of Nuclear Physics and Technology, Peking University, Beijing 100871, China}

\maketitle

\begin{abstract}
	Fragmentation cross section of $^{28}$Si + $^{9}$Be reaction at 75.8 MeV/u was analyzed for studying the decay mode of single-proton emission in $^{21}$Al (the proton-rich nucleus with neutron closed-shell of $N = 8$ and $T_z = -5/2$). With comparison between the measured fragmentation cross section and the theoretical cross section produced by EPAX3.1a for the observed nuclei (i.e. $^{20}$Mg, $^{21}$Al and $^{22}$Si), the expected yield for a particle stable $^{21}$Al was estimated. With exponential decay law, an upper limit of half-life of $13$ ns was determined. Using the single-proton penetration model, the upper limit of single-proton separation energy of $-105$ keV was deduced. This deduced mass limit agrees with the microscopic calculation based on nucleon-nucleon (NN) + three-nucleon (3N) forces in $sdf_{7/2}p_{3/2}$ valence space, which indicates the importance of 3N forces in $^{21}$Al.
\end{abstract}




\begin{multicols}{2}
Nuclear mass, one of the most important experimental observables, is a direct probe of fundamental interactions among participating nucleons. With the rapid development of accelerator over recent decades, experimentalists are able to investigate the nuclei far from the valley of stability, allowing us to study the exotic nuclei approaching the boundary of chart of nuclei. Three-nucleon (3N) force, firstly mentioned by Fujita and Miyazawa in 1957 \cite{3Norigin}, was discovered in the exotic region. Without including 3N forces, the double magicity of $^{48}$Ca and oxygen anormaly cannot be reproduced in the shell-model calculations derived from microscopic NN-only forces\cite{48Ca,24O,25O,26O}. This implies the 3N forces may play a crucial role in the exotic nuclei\cite{importance}. Up to now, only a few theoretical work regarding 3N forces have been conducted on the proton-rich nuclei, although they are crucial in nuclear structure, quantum many-body system and nuclear astrophysics. On the other hand, experimental studies near the proton drip line have been achieved\cite{1,2,3}, for instance recent study of $\beta$-delayed two-proton (2p) emission in the decay of $^{22}$Si\cite{22Si} and the $\beta$-delayed particle emission in $^{20}$Mg\cite{20Mg1}.

Recently, Holt et al.\cite{3N} started a theoretical work on proton-rich neutron closed-shell of $N = 8$ isotones, in which their ground-state energies relative to $^{16}$O, based on NN-only interaction and NN + 3N interaction in $sdf_{7/2}p_{3/2}$ valence spaces respectively, were calculated. These two microscopic calculations show a huge discrepancy for the nuclei $A > 19$ (i.e. $^{20}$Mg, $^{21}$Al and $^{22}$Si), which requires mass measurement data for evaluation. The mass measurement of $^{20}$Mg\cite{20Mg2} and $^{22}$Si\cite{22Si} has been done recently, and both results are in a good agreement with the calculation including 3N forces. For $^{21}$Al, no mass measurement has been achieved, although it is expected to be unbound (negative single-proton separation energy) by AME2003\cite{AME2003}, AME2012\cite{AME2012}, IMME\cite{IMME1} and ImKG\cite{ImKG}. Thus, the experimental mass information of $^{21}$Al is of importance to complete the systematic study of 3N forces along these proton-rich isotones.

In this article, the mass limit of $^{21}$ Al is deduced with the study of fragmentation cross section. The experiment was performed at the National Laboratory of Heavey Ion Research (HIRFL) of the Institude of Modern Physics, Lanzhou, China. A 75.8 MeV/nucleon primary beam $^{28}$Si with an average intensity of 37 enA accelerated by HIRFL cyclotrons impinged on an 1500 $\mu$m $^{9}$Be target. The projectile-like fragments were separated and purified by the first Radioactive Ion Beam Line in Lanzhou (RIBLL1)\cite{RIBLL1}, then the energy losses ($\Delta \mathrm{E}_1$ and $\Delta \mathrm{E}_2$) and the time of flight (ToF) were measured with two silicon detectors and two scintillation counters respectively\cite{TOFE}.
	
\begin{center}
\includegraphics[width=0.5\textwidth]{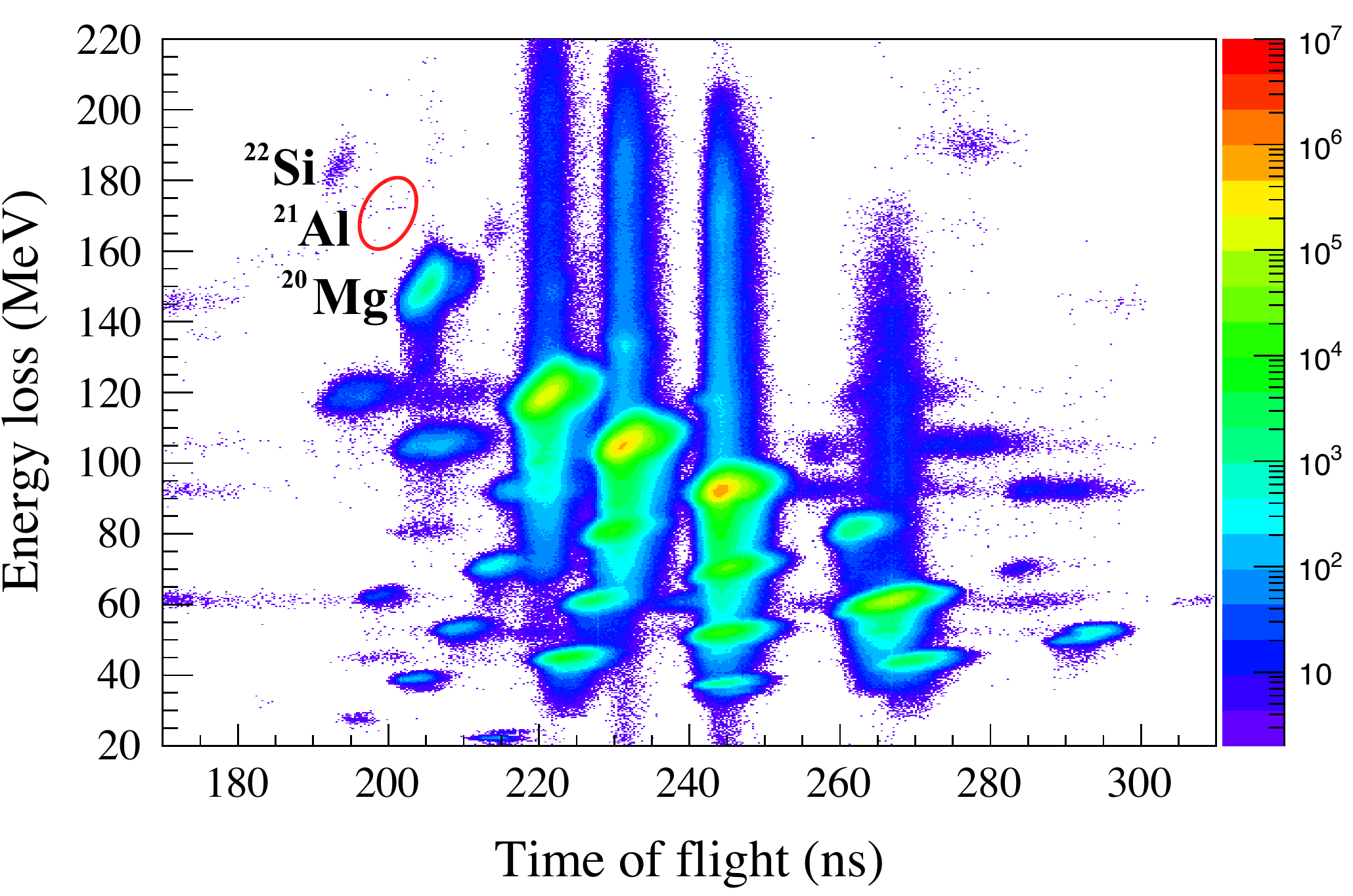}
\figcaption{\label{fig} The energy loss ($\Delta \mathrm{E}_1$) is plotted as a function of the time of flight (ToF) for all the events in the experiment. The red circle shows the expected region of $^{21}$Al.} 
\end{center}

\begin{center}
\tabcaption{ \label{tab1}  The measured event numbers, measured fragmentation cross section and theoretical cross section for the corresponding nuclei. The transmission efficiency was calculated by LISE\cite{LISE}.}
\footnotesize
\begin{tabular*}{80mm}{cccc}
\toprule Nucleus & Event numbers & $\sigma _\mathrm{measured}$ (mb) & $\sigma _\mathrm{theory}$ (mb)  \\
\hline
$^{22}$Si & $1.6 \times 10^3$  & $1.58 (35) \times 10^{-6}$ & $8.18 \times 10^{-6}$ \\
$^{20}$Mg & $3.4 \times 10^5$  & $1.03 (23) \times 10^{-3} $ & $2.40 \times 10^{-3}$ \\
$^{18}$Ne & $4.6 \times 10^7$  & $0.73 (16)$ & $ 0.67 $ \\

\bottomrule
\end{tabular*}
\vspace{0mm}
\end{center}
\vspace{0mm}

The Fig. \ref{fig} shows the two-dimensional particle identification plot of $\Delta \mathrm{E}_1$ versus ToF for the fragmentation reaction production. The measured event numbers, measured fragmentation cross section and theoretical cross section calculated by EPAX3.1a\cite{EPAX3.1a} are shown in Table \ref{tab1}. The measured fragmentation cross section and the theoretical cross section are consistent in the order of magnitude with each other, although the experimental value is smaller than theoretical value for $^{22}$Si and $^{20}$Mg, which could be explained by the inaccuracy of the EPAX calculation near the drip line.  Therefore, with the cross section in Table \ref{tab1} and EPAX3.1a cross section of $ \sigma \mathrm{(}^{21}\mathrm{Al)} = 1.4 \times 10^{-4} $ mb, the expected yield of $^{21}$Al in this measurement is estimated to be $2.1 \times 10^4$. The count in the region of $^{21}$Al in Fig. \ref{fig} is 59 and their energy losses in the second silicon detector $\Delta \mathrm{E}_2$ are scattered from 50 MeV to 260 MeV, which indicates that these events are the background since the energy loss distribution $\Delta \mathrm{E}_2$ of $^{21}$Al, if there is any, should be from 180 MeV to 200 MeV. The half-life limit can be calculated by assuming the isotopes of interest having decayed during the ToF with the order of 200 ns. By exponential decay law

\[
N(t) = (2.1 \times 10^4 )\left( \frac{1}{2} \right)^{\frac{t}{T_{1/2}}}
\]
and setting N(t) to be 1 (equivalent to no observation), the half-life of $^{21}$Al is calculated to be smaller than 13 ns. This half-life limit can be used to extract the upper bound single-proton separation energy based on the single-proton penetration model, in which the decay width is estimated by the multiplication of barrier penetration probability\cite{Sp1} and Wigner single-particle width\cite{Sp2,Sp22}. With an orbital angular momentum of $ l = 2 $ from the standard shell model, the single-proton separation energy of $^{21}$Al is estimated as $S_p < -105$ keV.  The mass excess of $^{21}$Al can be determined from the equation: $\Delta (^{21}\mathrm{Al}) = \Delta (^{20}\mathrm{Mg}) + \Delta (^{1}\mathrm{H}) - S_p$, where $\Delta (^{20}\mathrm{Mg}) = 17478$ keV\cite{20Mg2} and $\Delta (^{1}\mathrm{H}) = 7289$ keV \cite{AME2012}, giving $\Delta (^{21}\mathrm{Al}) > 24872$ keV. If the background is taken into account as the real events of $^{21}$Al, i.e. N(t) = 59, the half-life limit turns to be 23 ns and the corresponding single-proton separation energy limit is $-101$ keV which causes only negligible change in the mass excess limit.

\begin{center}
\includegraphics[width=0.5\textwidth]{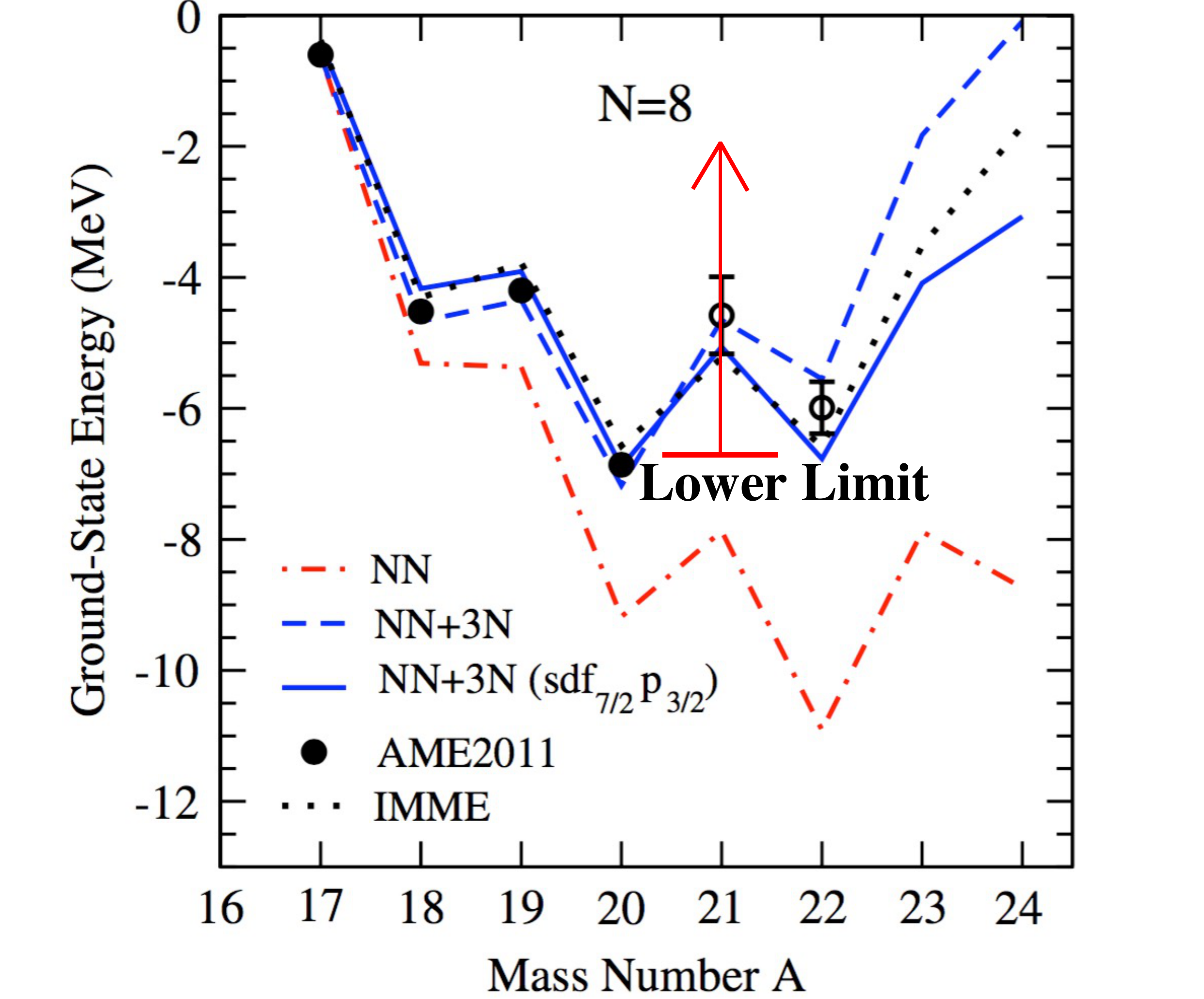}
\figcaption{\label{2} Adapted from Holt's paper\cite{3N}. Ground state energy of $N = 8$ isotones relative to $^{16}$O. AME2011 (extrapolation values are shown as open circles) and IMME values are shown. The NN-only results in the $sd$ shell and calculations based on NN + 3N forces in both $sd$ and $sdf_{7/2}p_{3/2}$ valence spaces are shown. The red arrow shows our deduced lower limit of ground state energy of $^{21}$Al relative to $^{16}$O.} 
\end{center}

Our deduced single-proton separation energy of $^{21}$Al, $S_p < -105$ keV, agrees with the calculation of isobaric multiplet mass equation (IMME), in which the mass excess is deduced from the mass excess of its mirror nuclide $\Delta (^{21}\mathrm{O}) = 8062$ (12) keV\cite{AME2012} through $\Delta (^{21}\mathrm{Al}) = \Delta (^{21}\mathrm{O}) - 2b(A,T)T_z$. With the global fit\cite{IMME1} and the readjustment fit\cite{IMME2} on b-coefficient respectively, the masses are deduced to be 25957 (12) keV and 25807 (12) keV. The corresponding single-proton separation energies $S_p$ are $-1191$ keV and $-1040$ keV respectively. Moreover, our result is consistent with the expected mass excess from AME2003 \textemdash  26120 (300) keV\cite{AME2003}, AME2012 \textemdash  26990 (400) keV\cite{AME2012} and ImKG \textemdash  26024 (12) keV\cite{ImKG}.

Our result also agrees with the many-body perturbation theory calculation for $^{21}$Al based on 3N + NN interaction in an extended $sdf_{7/2}p_{3/2}$ valence space, shown in Fig. \ref{2}. On the other hand, our result contradicts to the NN-only interaction calculation\cite{3N}, which indicates the existence of 3N forces.

With the study of fragmentation cross section, we determined the upper limit of half-life of $^{21}$Al ($T_{1/2} < 13 $ ns). Subsequently, the single-proton separation energy limit ($S_p < -105$ keV) is estimated based on the single-proton penetration model. Consequently, its mass excess is calculated to be $\Delta (^{21}\mathrm{Al}) > 24872$ keV, which is in agreement with calculation based on NN + 3N forces in $sdf_{7/2}p_{3/2}$ valence space. Our result indicates the importance of 3N forces, which should be included in the use of shell model.

\acknowledgments{We acknowledge the continuous effort of HIRFL operators for providing good-quality beams and ensuring compatibility of the electronics. This work was supported by the National Natural Science Foundation of China under Grant No. U1632136, No. U1432246, No. U1432127, No. 11375268, No. 11305272, No. 11475263, and the National Basic Research Program of China under Grant No. 2013CB834404.}
\end{multicols}

\vspace{15mm}
\vspace{-1mm}
\centerline{\rule{80mm}{0.1pt}}
\vspace{2mm}

\begin{multicols}{2}

\end{multicols}

\clearpage
\end{document}